\documentclass[12pt,a4paper]{article}
\usepackage{amsfonts,latexsym}
\usepackage{amsmath,amssymb}
\usepackage{graphicx,color}

\renewcommand{\thefootnote}{\fnsymbol{footnote}}

\begin{document}

\vspace{12mm}

\begin{center}
{{{\Large {\bf Mass-induced instability of SAdS black hole \\ in Einstein-Ricci cubic gravity }}}}\\[10mm]

{Yun Soo Myung\footnote{e-mail address: ysmyung@inje.ac.kr}}\\[8mm]

{Institute of Basic Sciences and Department  of Computer Simulation, Inje University Gimhae 50834, Korea\\[0pt]}

\end{center}
\vspace{2mm}

\begin{abstract}
We perform  the stability analysis of Schwarzschild-AdS (SAdS) black
hole in the Einstein-Ricci cubic  gravity. It shows that the
Ricci tensor perturbations exhibit unstable modes
for small black holes. We call this the mass-induced instability of SAdS black hole because the instability of small
black holes  arises from  the
massiveness in the linearized Einstein-Ricci cubic gravity,  but not a feature of higher-order derivative theory
giving ghost states.  Also, we point out that the correlated stability conjecture holds for the SAdS black hole
by computing the Wald entropy of SAdS black hole in Einstein-Ricci cubic gravity.
\end{abstract}
\vspace{5mm}

\vspace{1.5cm}

\hspace{11.5cm}
\newpage
\renewcommand{\thefootnote}{\arabic{footnote}}
\setcounter{footnote}{0}

%%%% Introduction %%%%

\section{Introduction}

The study of higher-derivative gravity theories has attracted critical attention in quantum  gravity.
Stelle's seminal work~\cite{Stelle:1976gc} has shown that  the fourth-order gravity  is renormalizable and  has a finite Newtonian potential at origin.
However, this gravity  belongs to a nonunitary theory because it has a massive spin-2 pole with negative residue which could be interpreted   as a state of negative norm (ghost).
It turns out that the infinite derivative gravity (non-local gravity) is ghost-free and renormalizable around the Minkowski spacetime background when one chooses the exponential form of an  entire function
~\cite{Modesto:2011kw,Biswas:2011ar}. The most general class of theories that are ghost-free on any background is Lovelock gravity~\cite{Lovelock:1971yv} whose terms of order $k$ in the curvature are topological in $d=2k$ dimensions and vanish identically for $d<2k$.  Quasi-topological gravities provide additional example in higher dimensions than four~\cite{Myers:2010ru,Oliva:2010eb,Cisterna:2017umf}, but all quasi-topological theories are trivial in four dimensions.

Recently, the Einsteinian cubic gravity of ${\cal L}_{\rm EC}=R-2\Lambda_0-\lambda {\cal P}/6$  with ${\cal P}$ Riemann polynomials was introduced to indicate that it is neither topological nor trivial in four dimensions~\cite{Bueno:2016xff}.
It was shown that black hole solutions of this gravity have a number of interesting properties~\cite{Hennigar:2017ego,Ahmed:2017jod,Bueno:2016lrh}, but these belong to either  numerical or approximate solutions.
That is, the Einstein equation  cannot be solved analytically. An obstacle to studying these black holes is the lack of an analytic solution.
At this stage, we remind the reader that at the critical points, the Einsteinian cubic gravity admits AdS black boles in four and five dimensions~\cite{Feng:2017tev}.

On the other hand, it is important to note that  Ricci  polynomials are much more manageable, compared to Riemann polynomials. The Ricci cubic gravity~\cite{Li:2017ncu} can be composed of three terms from six cubic invariants in four dimensions, which has a similar property to the Ricci quadratic gravity (fourth-order gravity). It is known that the linearized theory of any higher-order gravity around a maximally symmetric background can be mapped into the linearized theory of fourth-order gravity~\cite{Bueno:2016ypa}. This may imply that if one performs the linear stability analysis for a black hole obtained from a general quadratic gravity, these results could  apply  to analyzing the stability of the same black hole obtained from any higher-order gravity.
A crucial benefit of Ricci cubic  gravity is that the Schwarzschild black hole to Einstein gravity is a solution to this theory.
In this case,  the solution represented by mass $r_0=2M$  describes the gravitational field outside of a static matter distribution, because its linearized theory reduces to that of Einstein gravity. It was proved that the  only theories susceptible of admitting solutions with $g_{tt}g_{rr}=-1$ and representing the exterior field of a spherically symmetric distribution of mass are those that only propagate a massless spin-2 mode with 2 DOF (degrees of freedom) on the vacuum~\cite{Bueno:2017sui}.
We note that  the  Schwarzschild-AdS (SAdS) black hole to Einstein gravity with a cosmological constant is  a solution to the theory, but the SAdS is not a solution to the Einsteinian cubic gravity.  Importantly, one can construct a covariant linearized gravity on the SAdS black hole in Ricci cubic gravity, but the covariant linearized theory
 of Riemann  polynomials is allowed only on a maximally symmetric vacuum of  AdS$_4$ spacetimes.
However, the SAdS solution does not describe the exterior field of a spherically symmetric distribution of mass in Einstein-Ricci cubic gravity because its linearized equation (\ref{LE-5}) describing 5 DOF  could not reduce to that of Einstein gravity.
  
  Also, the (in)stability of SAdS black hole is known for fourth-order gravity~\cite{Myung:2013bow} and thus, this result will be compared to that of Ricci cubic gravity.    These are  a few of reasons why we wish to introduce the Ricci cubic gravity as a higher-order gravity in the study of a black hole.

If a SAdS black hole is obtained  from the Ricci cubic gravity, one has to ask what this all  mean for ``physical black hole"?
A physical black hole can be selected by the stability analysis.
If it is stable against the metric perturbation, one accepts it as a physical black hole.
If not, one has to reject it. This is a long-standing issue since 1957.
First of all, the linearized equation around the
Schwarzschild black hole is given by $\delta R_{\mu\nu}(h)=0$ in Einstein gravity. Then, the metric perturbation
$h_{\mu\nu}$ is classified depending on the transformation
property under parity, namely odd  and even. Using the
Regge-Wheeler~\cite{Regge:1957td} and Zerilli gauge~\cite{Zerilli:1970se}, one
obtains two distinct perturbations: odd and even parities. It turns out that the Schwarzschild black
hole is stable against the metric perturbation~\cite{Vishveshwara:1970cc,chan}.
Investigating the stability analysis of the SAdS black hole in Einstein gravity with a cosmological constant, one might  use the linearized Einstein equation $\delta G_{\mu\nu}(h)=0$.
It turns  out to be stable by following the Regge-Wheeler
prescription~\cite{Cardoso:2001bb,Ishibashi:2003ap}.

We would like to mention that the Regge-Wheeler
prescription is limited to the second-order gravity.
Thus, one could not implement the Regge-Wheeler prescription to perform the stability analysis of black hole found in higher-order gravity.
For a higher-order gravity, Whitt~\cite{Whitt:1985ki} has argued  that provided both massive spin-0 and spin-2 gravitons
are non-tachyonic, the Schwarzschild black hole is classically
stable in  fourth-order gravity  when using  the
linearized Ricci tensor equation. In this case, the linearized Ricci tensor could represent a massive spin-2 field.
 Considering an auxiliary field formulation for decreasing a fourth-order gravity to a second-order theory of gravity~\cite{Mauro:2015waa},
one found that the linearized equation for Ricci tensor $\delta R_{\mu\nu}$ is transformed exactly into that for  auxiliary field $\psi_{\mu\nu}$.
Hence, one does not worry
about the ghost (an unhealthy massive spin-2 field) problem  arising from the fourth-order gravity
 because the linearized Ricci tensor  $\delta R_{\mu\nu}$ as  a healthy  massive spin-2 field  satisfies  a second-order equation~\cite{Stelle:2017bdu}.
Visiting  this stability issue again, it has  shown that
the small  black hole in Einstein-Weyl gravity  is unstable against $s(l=0)$-mode Ricci tensor perturbation, while the large black hole is stable
against $s$-mode perturbation~\cite{Myung:2013doa}.  Actually, this was
performed by comparing the linearized Ricci tensor equation  with
the linearized metric equation around the five-dimensional black string where the Gregory-Laflamme instability appeared~\cite{Gregory:1993vy}.

In addition, one observes that there was a close connection between thermodynamic instability and
classical [Gregory-Laflamme] instability for the black strings/branes.  This Gubser-
Mitra proposal~\cite{Gubser:2000ec} was referred to as the correlated stability conjecture (CSC)~\cite{Harmark:2007md}.
The CSC states that the classical instability of a black string/brane with translational symmetry
and infinite extent sets in precisely, when the corresponding thermodynamic system
becomes  thermodynamically unstable (that is, either Hessian matrix1 $<0$ or heat capacity $< 0$).
Here the additional assumption of translational symmetry and infinite extent has been
added to ensure that finite size effects do not spoil the thermodynamic nature of the argument
and to exclude a well-known case of the Schwarzschild black hole. A famous example of holding in the CSC is the five-dimensional black string.
It is known that the Schwarzschild  black hole is classically
stable, but thermodynamically unstable because of its negative heat capacity. Also, the SAdS black hole is stable against the metric perturbation, whereas
the small (large) black hole with $r_+<r_*=\ell/\sqrt{3}(r_+>r_*)$  is thermodynamically unstable (stable) because of negative (positive) heat capacity.
Therefore, the last two examples show violation of the CSC and inapplicability of $s$-mode perturbation because a massless spin-2 mode perturbation starts from $l=2$.

However, the situation is changed when one analyzes   a black hole found in a fourth-order gravity (a massive gravity) where the infinite extent with translational symmetry is absent clearly. An important thing being different from the Einstein gravity  is the appearance  of a massive spin-2 mode.
A massive spin-2 mode allows  us  to define a mass squared $M^2$ and to analyze the black hole with $s(l=0)$-mode perturbation.
Considering a  setting  $e^{\frac{\Omega}{r_0} t} e^{i\frac{k}{r_0} z}$ for a black string perturbation~\cite{Gregory:1993vy}, there exists a critical wave number $k^c$ where
for $k<k^c(k>k^c)$, the black sting is unstable (stable) against metric perturbations. There is an unstable (stable) mode for any wavelength large (smaller)
than the critical wavelength $\lambda_{\rm GL}=\frac{2\pi r_0}{k^c}$.
Here, the mass  $M$ of a massive spin-2 mode  plays a role of  $k/r_0$ because the mass operator is given by $\partial_z$. This implies that the massiveness ($M^2\not=0$)
takes over a black string located in $z$ direction effectively. The Gregory-Laflamme instability is an $s$-wave unstable mode from the four-dimensional
perspective~\cite{Stelle:2017bdu}.
In this respect, the dRGT massive gravity~\cite{deRham:2010kj} having a Schwarzschild solution is subject directly to an $s$-wave instability~\cite{Babichev:2013una,Brito:2013wya}. Here, we pay our attention to an equivalence between 4D black hole in linearized massive gravity and 5D black string in linearized  Einstein gravity from the four-dimensional perspective.

Furthermore, the Gregory-Laflamme instability condition (massiveness) picks up the small AdS black hole with $r_+<r_*$ which is thermodynamically unstable in fourth-order gravity. It is known that
 the CSC holds for the SAdS black hole in Einstein-Weyl gravity   by establishing a connection between the thermodynamic instability  and the Gregory-Laflamme instability~\cite{Myung:2013uka}. Also, the CSC holds for the BTZ black hole regardless of the horizon radius $r_+$ in
 three-dimensional new massive gravity.
 Hence it is quite interesting to check whether the CSC  holds for a black hole found in higher-order gravity theory.

In this work, we will investigate  classical instability and thermodynamics of SAdS black
holes in Einstein-Ricci cubic  gravity. Hereafter, we rename Ricci cubic  gravity as Einstein-Ricci cubic gravity for a precise definition.
We perform the stability analysis of SAdS black hole in Einstein-Ricci cubic gravity by introducing the Gregory-Laflamme scheme.
Computing the Wald entropy, we derive  other thermodynamic quantities by making use of  the first-law of thermodynamics in Einstein-Ricci cubic  gravity.
  Finally, we wish to establish a connection between  the Gregory-Laflamme instability and the thermodynamic
instability of SAdS black holes in $\alpha=-3\beta$ Einstein-Ricci cubic gravity.  This will provide  another example for which the CSC holds.
\section{Einstein-Ricci cubic gravity}
We start with the Einstein-Ricci cubic  gravity (ER) in four dimensional  spacetimes~\cite{Li:2017ncu}
\begin{eqnarray}
S_{\rm ER}&\equiv&\frac{1}{16 \pi}\int d^4 x\sqrt{-g} {\cal L}_{\rm ER}\nonumber \\
&=&\frac{1}{16 \pi}\int d^4 x\sqrt{-g} \Big[\kappa(R-2\Lambda_0)
+(e_1R^2+e_2  R_{\mu\nu}R^{\mu\nu})R+e_3 R^\mu_\nu R^\nu_\rho R^\rho_\mu\Big] \label{Action}
\end{eqnarray}
 with $\kappa=1/G$ the inverse of Newtonian constant, $\Lambda_0$ the bare cosmological constant, and $(e_1,e_2,e_3)$ three cubic  parameters.
Here we observe  from the second term of (\ref{Action}) that the fourth-order gravity is embedded into the Einstein-Ricci cubic gravity.
From the action  (\ref{Action}),  the Einstein equation is derived to be
\begin{eqnarray}
P_{\mu\alpha\beta\gamma}R_\nu~^{\alpha\beta\gamma}-\frac{1}{2}g_{\mu\nu}{\cal L}_{\rm ER}
-2\nabla^\alpha\nabla^\beta P_{\mu\alpha\beta\nu}=0,  \label{equa1}
\end{eqnarray}
where the $P$-tensor  is defined  by
\begin{equation}
P_{\mu\nu\rho\sigma}=\frac{\partial {\cal L}_{\rm ER}}{\partial R^{\mu\nu\rho\sigma}}.
\end{equation}
Explicitly, it  takes the form
\begin{eqnarray}
P_{\mu\nu\rho\sigma}&=& \frac{\kappa}{2} (g_{\mu\rho}g_{\nu\sigma}-g_{\mu\sigma}g_{\nu\rho})+\frac{3e_1}{2}R^2(g_{\mu\rho}g_{\nu\sigma}-g_{\mu\sigma}g_{\nu\rho})\nonumber \\
&+&\frac{e_2}{2}R_{\alpha\beta}R^{\alpha\beta}(g_{\mu\rho}g_{\nu\sigma}-g_{\mu\sigma}g_{\nu\rho})+\frac{e_2}{2}R(g_{\mu\rho}R_{\nu\sigma}-g_{\mu\sigma}R_{\nu\rho}-g_{\nu\rho}R_{\mu\sigma}+g_{\nu\sigma}R_{\mu\rho})\nonumber \\
&+& \frac{3e_3}{4}(g_{\mu\rho}R_{\nu\gamma}R^\gamma_\sigma-g_{\mu\sigma}R_{\nu\gamma}R^\gamma_\rho-g_{\nu\rho}R_{\mu\gamma}R^\gamma_\sigma+g_{\nu\sigma}R_{\mu\gamma}R^\gamma_\rho).  \label{equa2}
\end{eqnarray}
At this stage, we propose that  a SAdS black
hole solution to the Einstein gravity,
 \begin{equation} \label{sch}
ds^2_{\rm SAdS}=\bar{g}_{\mu\nu}dx^\mu
dx^\nu=-f(r)dt^2+\frac{dr^2}{f(r)}+r^2d\Omega^2_2
\end{equation}
 is also a solution to Eq.(\ref{equa1}).
Here the metric function is given by
 \begin{equation} \label{num}
f(r)=1-\frac{r_0}{r}-\frac{\Lambda}{3}r^2,~~\Lambda=-\frac{3}{\ell^2}
\end{equation}
with $\ell$  the
curvature radius of AdS$_4$ spacetimes.
The effective cosmological constant $\Lambda$ is related to the bare cosmological constant as
\begin{equation}
\kappa \Lambda -(16e_1+4e_2+e_3)\Lambda^3=\kappa \Lambda_0.
\end{equation}
We note that a black hole mass
parameter is determined as
\begin{equation}
r_0=r_+\Big(1+\frac{r_+^2}{\ell^2}\Big)
\end{equation}
which  is not surely  the horizon radius
$r_+$. Hereafter we
denote all background quantities with the ``overbar''. The background Ricci tensor and Ricci scalar  are given by
\begin{equation} \label{beeq}
\bar{R}_{\mu\nu}=\Lambda\bar{g}_{\mu\nu},~\bar{R}=4\Lambda.
\end{equation}
In this case, one notes that the background
 $P$-tensor takes a maximally symmetric form
\begin{equation} \label{p-tensor}
\bar{P}_{\mu\nu\rho\sigma}=\frac{1}{2}\Big(\kappa+48e_1\Lambda^2+12e_2\Lambda^2+3\Lambda^2e_3\Big)(\bar{g}_{\mu\rho}\bar{g}_{\nu\sigma}-\bar{g}_{\mu\sigma}\bar{g}_{\nu\rho}).
\end{equation}
It is easy to show that the SAdS black hole  (\ref{sch})
to the Einstein equation of $G_{\mu\nu}-\Lambda\bar{g}_{\mu\nu}=0$  is also the solution
to the Einstein-Ricci cubic gravity when one substitutes (\ref{p-tensor})  together with (\ref{beeq}) into (\ref{equa1}).
However, it is important to note  that the background Riemann tensor for the SAdS black hole is not given by the  AdS$_4$-curvature tensor
\begin{equation} \label{ads-rt}
\bar{R}_{\mu\nu\rho\sigma}\not=\bar{R}_{\mu\nu\rho\sigma}^{\rm AdS_4}=\frac{\Lambda}{3}(\bar{g}_{\mu\rho}\bar{g}_{\nu\sigma}-\bar{g}_{\mu\sigma}\bar{g}_{\nu\rho}),
\end{equation}
which means that the SAdS spacetimes   (\ref{sch}) is not a maximally symmetric vacuum.

\section{Linearized Einstein-Ricci cubic gravity}
 To perform
the stability analysis, we introduce the metric
perturbation around the SAdS black hole as
\begin{eqnarray} \label{m-p}
g_{\mu\nu}=\bar{g}_{\mu\nu}+h_{\mu\nu}.
\end{eqnarray}
Then, we may define the linearized  Ricci tensor and scalar  as
\begin{equation}
\delta \tilde{R}_{\mu\nu}=\delta R_{\mu\nu}-\Lambda h_{\mu\nu},~\delta R=\delta(g^{\mu\nu}R_{\mu\nu})=\bar{g}^{\mu\nu}\delta \tilde{R}_{\mu\nu},
\end{equation}
where
\begin{eqnarray}
\delta R_{\mu\nu}&=&\frac{1}{2}\Big(\bar{\nabla}^{\rho}\bar{\nabla}_{\mu}h_{\nu\rho}+
\bar{\nabla}^{\rho}\bar{\nabla}_{\nu}h_{\mu\rho}-\bar{\nabla}^2h_{\mu\nu}-\bar{\nabla}_{\mu}
\bar{\nabla}_{\nu}h\Big), \label{ricc-t} \\
\delta R&=& \bar{g}^{\mu\nu}\delta
R_{\mu\nu}-h^{\mu\nu}\bar{R}_{\mu\nu}= \bar{\nabla}^\mu
\bar{\nabla}^\nu h_{\mu\nu}-\bar{\nabla}^2 h-\Lambda h
\label{Ricc-s}.
\end{eqnarray}
with $h=h^\rho_\rho$. In this case, the linearized Einstein tensor can be written  by
\begin{equation}
\delta G_{\mu\nu}=\delta \tilde{R}_{\mu\nu}-\frac{1}{2}\bar{g}_{\mu\nu}\delta R. \label{LET}
\end{equation}
Introducing two new parameters  $\alpha=4e_2+3e_3$ and $\beta=2(6e_1+e_2)$, the linearized Einstein equation can be rewritten  compactly as
\begin{eqnarray}
&&\kappa \delta G_{\mu\nu}+\Lambda^2(3\alpha+4\beta)\delta \tilde{R}_{\mu\nu}
-\frac{1}{2}\Lambda^2 \alpha\bar{g}_{\mu\nu} \delta R \nonumber \\
&&-\Lambda\alpha\bar{\Delta}_{\rm L}\delta G_{\mu\nu}-\Lambda (\alpha+2\beta)(\bar{\nabla}_\mu \bar{\nabla}_\nu-\bar{g}_{\mu\nu}\bar{\square} )\delta R=0, \label{l-eineq}
\end{eqnarray}
where the background Lichnerowicz operators are defined  by acting  on scalar and tensor, respectively,
\begin{eqnarray}
&&\bar{\Delta}_{\rm L}\delta R=-\bar{\square}\delta R, \nonumber \\
&& \bar{\Delta}_{\rm L} \delta \tilde{R}_{\mu\nu}=-\bar{\square}\delta \tilde{R}_{\mu\nu}-2 \bar{R} _{\mu\rho\nu \sigma}\delta \tilde{R}^{\rho\sigma}+\bar{R}_\mu^\rho \delta \tilde{R}_{\rho\nu}
+\bar{R}_\nu^\rho \delta \tilde{R}_{\rho\mu}.
\end{eqnarray}
The linearized equation (\ref{l-eineq}) is a second-order equation for $\delta G_{\mu\nu}$ and $\delta R$, but it becomes a fourth-order equation for $h_{\mu\nu}$.
This is surely  our expectation that the linearized theory of any higher-order gravity is a fourth-order theory of gravity.
Here, we have a fourth-order equation (\ref{l-eineq})  because of the inclusion of the cosmological constant $\Lambda_0$ in the action (\ref{Action}).
Putting $\Lambda=0(\Lambda_0=0)$ yields Ricci-flat spacetimes on which cubic curvature terms give no contributions to the linearized equations, leading to
$\delta R_{\mu\nu}=0$.  This is one reason why we included the  cosmological constant in the beginning action (\ref{Action}),  compared to the fourth-order gravity.
In addition, it is worth noting  that  Eq.(17) leads to (2.15) in Ref.\cite{Bueno:2016ypa} when replacing $\bar{R}_{\mu\nu\rho\sigma}$ by $\bar{R}_{\mu\nu\rho\sigma}^{\rm AdS_4}$ in (\ref{ads-rt}).

Taking the trace of Eq.(\ref{l-eineq}) leads to the linearized Ricci scalar equation
\begin{equation}
2\Lambda(\alpha+3\beta)\bar{\square}\delta R+[-\kappa+\Lambda^2(\alpha+4\beta)] \delta R=0.\label{LE-0}
\end{equation}
One notes that Eq.(\ref{l-eineq}) is a coupled second-order equation for $\delta \tilde{R}_{\mu\nu}$ and $\delta R$, which seems to be difficult to be solved.
One way to avoid this difficulty is to split Eq.(\ref{l-eineq}) into the traceless and trace parts by choosing $\alpha$ and $\beta$  appropriately.
For this purpose, we introduce a traceless Ricci tensor as
\begin{equation}
 \delta\hat{ R}_{\mu\nu}= \delta \tilde{R}_{\mu\nu}-\frac{1}{4}\bar{g}_{\mu\nu} \delta R,~~ \delta \hat{R}=0.
\end{equation}
Then, Eqs.(\ref{l-eineq}) and (\ref{LE-0}) lead to
\begin{eqnarray}
&&\Lambda \alpha (\bar{\Delta}_{\rm L} -2\Lambda+\mu^2_2)\delta \hat{ R}_{\mu\nu}=-\Lambda (\alpha+2\beta) \Big(\bar{\nabla}_\mu \bar{\nabla}_\nu-\frac{1}{4}\bar{g}_{\mu\nu}\bar{\square}\Big)\delta R,\label{LE-1}\\
&&2\Lambda(\alpha+3\beta)(\bar{\square}-\mu_0^2) \delta R=0, \label{LE-2}
\end{eqnarray}
where the  mass squared $\mu^2_2$ for spin-2 mode,  and the mass squared $\mu^2_0$  for  spin-0 mode  are given by
\begin{eqnarray}
 \mu_2^2&=&2\Lambda-\frac{(3\alpha+4\beta)\Lambda^2+\kappa}{\Lambda \alpha},~~\mu^2_0=\frac{\kappa-(\alpha+4\beta)\Lambda^2}{2\Lambda(\alpha+3\beta)}. \label{masses}
\end{eqnarray}
First of all, decoupling of all massive modes requires either $\Lambda=0$ or $ \alpha=\beta=0$.  The former case corresponds to the Ricci-flat spacetimes on which cubic curvature tensor gives no
 contribution to the linearized Ricci tensor equation.
On the other hand, the latter case yields the quasi-topological Ricci cubic gravity whose linearized equation is exactly the same form as in the Einstein gravity with a cosmological constant~\cite{Li:2017ncu}.
Hence, these cases lead to a  linearized second-order gravity.

A nontrivial decoupling between traceless and trace parts may occur when choosing a condition of  $\alpha=-2\beta$. This case corresponds to the linearized fourth-order (Einstein-Ricci quadratic)  gravity
~\cite{Lu:2017kzi} because one parameter $e_3$ can be represented by $e_1$ and $e_2$.
In this case, Eqs.(\ref{LE-1}) and (\ref{LE-2}) lead to the massive spin-2 and massive spin-0 equations, separately,
\begin{eqnarray}
&&(\bar{\Delta}_{\rm L} -2\Lambda+\mu^2_2)\delta \hat{ R}_{\mu\nu}=0, \label{LE-3} \\
&&(\bar{\square}-\mu_0^2) \delta R=0. \label{LE-4}
\end{eqnarray}
Here the mass squared $\mu^2_2$ and $\mu^2_0$ are given by
\begin{equation}
\mu^2_2=\frac{\kappa\ell^2}{3\alpha}-\frac{3}{\ell^2},~\mu^2_0=\frac{\kappa\ell^2}{3\alpha}+\frac{3}{\ell^2}.
\end{equation}
However, the number of  degrees of freedom (DOF) for $\delta \hat{ R}_{\mu\nu}$ cannot be reduced to five because the contracted Bianchi identity of $\bar{\nabla}^\mu\delta G_{\mu\nu}=0$ does not imply
the transverse condition,
\begin{equation}
\bar{\nabla}^\mu\delta \hat{R}_{\mu\nu}=\frac{1}{4}\bar{g}_{\mu\nu}\bar{\nabla}_\nu \delta R \nrightarrow \bar{\nabla}^\mu\delta \hat{R}_{\mu\nu}=0
\end{equation}
due to $\delta R\not=0$ in  $\alpha=-2\beta$ Einstein-Ricci cubic gravity.

A promising choice may be  done  by requiring the non-propagation of the Ricci scalar.
From Eq.(\ref{LE-0}), imposing the condition of $\alpha=-3\beta$, we have a constraint of  non-propagating Ricci scalar
\begin{equation}
\delta R=0. \label{MLE-0}
\end{equation}
Also, one finds from (\ref{masses}) that the  mass squared $\mu^2_0$ of  massive spin-0 blows up, which means that the massive spin-0 is decoupled from the theory.
This  case corresponds to the linearized Einstein-Weyl gravity~\cite{Myung:2013bow,Myung:2013uka} because the linearized Ricci scalar is decoupled from the theory.
Considering $\delta G_{\mu\nu}$ in (\ref{LET}) together with $\delta R=0$, Eq.(\ref{l-eineq}) leads to
the massive spin-2 equation for $\delta \hat{R}_{\mu\nu}$ as
\begin{equation}
(\bar{\Delta}_{\rm L} -2\Lambda+M^2_2)\delta \hat{ R}_{\mu\nu}=0 \label{LE-5}
\end{equation}
with the mass squared
\begin{equation}
M^2_2=2\Lambda-\frac{(3\alpha+4\beta)\Lambda^2+\kappa}{\Lambda \alpha}|_{\beta=-\alpha/3}=\frac{\kappa\ell^2}{3\alpha}-\frac{1}{\ell^2}. \label{MLE-1}
\end{equation}
Here, one requires $M^2_2>0$ to avoid the tachyonic instability of $\delta \hat{ R}_{\mu\nu}$ propagating on the SAdS black hole~\cite{Liu:2011kf}.
Taking into account Eq.(\ref{MLE-0}), the contracted Bianchi identity provides a desired  transverse condition
\begin{equation}
\bar{\nabla}^\mu\delta \hat{R}_{\mu\nu}=0. \label{MLE-3}
\end{equation}
Hence, the DOF of $\delta \hat{R}_{\mu\nu}$ becomes five from the counting of  $10-4-1=5$ in  $\alpha=-3\beta$ Einstein-Ricci cubic gravity.

\section{Black hole stability}
\subsection{SAdS black hole in Einstein gravity}
First of all, in $\Lambda_0=0$ Einstein-Ricci cubic gravity (Einstein gravity), the linearized equation around the
Schwarzschild black hole is given by $\delta R_{\mu\nu}(h)=0$
with $\delta R_{\mu\nu}(h)$ in  (\ref{ricc-t}). Then, the metric perturbation
$h_{\mu\nu}$ is classified depending on the transformation
properties under parity, namely odd  and even. Using the
Regge-Wheeler~\cite{Regge:1957td} and Zerilli gauge~\cite{Zerilli:1970se}, one
obtains two distinct perturbations: odd with 2 DOF and even with 4
DOF. This implies that even though one starts with 6 DOF  under  the
Regge-Wheleer gauge, the propagating DOF   is two
for a  massless spin-2 metric tensor  $h_{\mu\nu}$.  It turned out that the Schwarzschild black
hole is stable against the metric perturbation~\cite{Vishveshwara:1970cc,chan}.

Performing the stability analysis of the SAdS black hole in $\alpha=\beta=0$ Einstein-Ricci cubic
gravity (quasi-topological Ricci cubic gravity)~\cite{Li:2017ncu}, one might  use the linearized equation $\delta G_{\mu\nu}(h)=0$  with $\delta G_{\mu\nu}(h)$ in (\ref{LET}).
It turns  out to be stable by following the Regge-Wheeler
prescription~\cite{Cardoso:2001bb,Ishibashi:2003ap,Moon:2011sz}. In
these cases, the $s(l=0)$-mode analysis is not necessary to show the
stability of the Schwarzschild  and SAdS black holes because the
massless spin-2 mode starts from $l=2$.

%%%%%%%%%%%%%%%%%%%%%%%%%%%%%%%%%%%%%%%%%%%%%%%%%%%%%%%%%%%%%%%%%%%%%%%%%
\subsection{SAdS black hole in Einstein-Weyl gravity}
The Regge-Wheeler prescription is no longer suitable for performing the stability analysis of the SAdS black hole in fourth-order gravity,
 since it focused on a linearized second-order gravity.
One may  explore the black hole stability by means of the second-order equation for the linearized Ricci tensor $\delta R_{\mu\nu}$,
  instead of the fourth-order equation for the metric perturbation $h_{\mu\nu}$~\cite{Whitt:1985ki,Stelle:2017bdu}.
 The  $s$-mode  analysis is an essential tool  to detect  the instability of
small SAdS black holes~\cite{Myung:2013doa} obtained from
Einstein-Weyl gravity given by
\begin{equation} \label{EWG}
{\cal L}_{\rm EW}=\kappa(R-2\Lambda)-\frac{3b}{2}C_{\mu\nu\rho\sigma}C^{\mu\nu\rho\sigma}
\end{equation}
with $C_{\mu\nu\rho\sigma}$ the Weyl tensor.
Considering  thermodynamics of  the SAdS black hole in (\ref{sch}), we usually denote the small (large) SAdS black holes by the condition of $r_+<r_*(r_+>r_*)$ with $r_*=\ell/\sqrt{3}$ and  $\ell^2=-3/\Lambda$.
Here $r_*$ is a position where the heat capacity blows up [see Fig. 4].
Its linearized equation is given by
\begin{equation} \label{EWL-eq}
(\bar{\Delta}_{\rm L}-2\Lambda+M^2)\delta G_{\mu\nu}(h)=0,
\end{equation}
with  the mass squared of massive spin-2 mode
\begin{equation}
 M^2=\frac{\kappa}{3b}+\frac{2\Lambda}{3}=\frac{\kappa}{3b}-\frac{2}{\ell^2}.
\end{equation}
Here, one requires $M^2>0$ to avoid the tachyonic instability of $\delta G_{\mu\nu}$ propagating on the SAdS black hole background~\cite{Liu:2011kf}.
We depict $M$ as a function of $b$ with $\kappa=1$ in Fig. 1. It is worth noting that $M^2$ is zero at $b=b^*=\ell^2/6=16.6$, which corresponds to the critical gravity.  The tachyon-free condition implies the allowed range for $b$ as
\begin{equation}
0<b<b^*.
\end{equation}
However, $M^2>0$ is not a necessary and sufficient condition to guarantee a stable SAdS black hole.
If $M^2<0(b>b^*)$, one does not need to consider a further analysis because it implies the tachyonic instability.
\begin{figure*}[t!]
   \centering
   \includegraphics{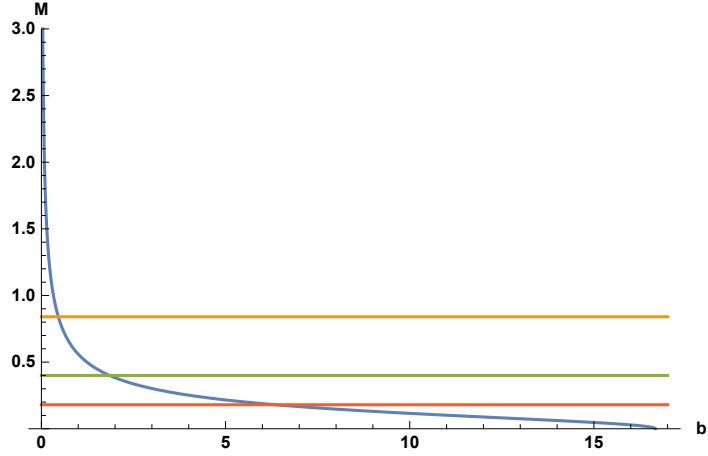}
\caption{Plots of mass $M$ for a massive spin-2 as a function of $b$  with  $l=10$ and $\kappa=1$ in the linearized  Einstein-Weyl gravity.
The $M=0(b=b^*)$ case corresponds to the critical  gravity, while three horizontal lines denote $M^c={\cal O}(1)/r_0|_{r_+=1,2,4}=0.84,0.40,0.18$.
}
\end{figure*}

Here, one has  $\delta R=0$ which means that there is no  massive spin-0 (Ricci scalar) propagation in Einstein-Weyl gravity.
Therefore, the linearized Einstein tensor $\delta G_{\mu\nu}$ satisfies the transverse-traceless condition of $\bar{\nabla}^\mu \delta G_{\mu\nu}=0$ and $\delta G=-\delta R=0$.
Hence, its DOF is five from counting of $10-1-4=5$.
The even-parity metric perturbation in Einstein gravity is
used  for a single $s$-mode analysis in the Einstein-Weyl gravity.
Its form is given by $\delta G_{tt},~\delta G_{tr},~\delta G_{rr},$ and $\delta G_{\theta\theta}$ as displayed in the matrix form
\begin{eqnarray}
\delta G^e_{\mu\nu}=e^{\Omega t} \left(
\begin{array}{cccc}
\delta G_{tt}(r) & \delta G_{tr}(r) & 0 & 0 \cr \delta G_{tr}(r) &\delta  G_{rr}(r) & 0 & 0 \cr
0 & 0 & \delta G_{\theta\theta}(r) & 0 \cr 0 & 0 & 0 & \sin^2\theta \delta G_{\theta\theta}(r)
\end{array}
\right). \label{evenp}
\end{eqnarray}
Even though one starts with 4 DOF, they are  related to each other
when one uses the Bianchi identity of $\bar{\nabla}^\mu\delta G_{\mu\nu}=0$ together  with $\delta G=-\delta R=0$.
\begin{figure*}[t!]
   \centering
   \includegraphics{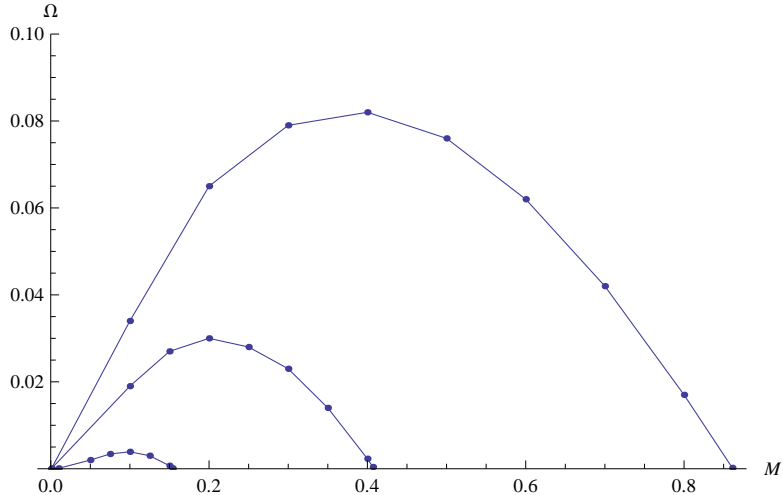}
\caption{Plots of unstable modes ($\bullet$) on three curves for small black holes $r_+=1,2,4<r_*=5.7$
with  $l=10$. The $y$-axis denotes $\Omega$ in $e^{\Omega t}$, while  the $x$-axis  is the  mass $M$. The smallest curve
represents $r_+=4$, the medium denotes $r_+=2$, and the largest one
shows $r_+=1$.  }
\end{figure*}
%%%%%%%%%%%%%%%%%%%%%%%%%%%%%%%%%%%%%%%%%%%%%%%%%%%%%%%%%%%%%%%%%%%%%%%%%
Hence, we  derive one
decoupled second-order equation for  $\delta G_{tr}$,
\begin{equation} \label{secondG-eq} A(r;r_0,\ell,\Omega^2,M^2)
\frac{d^2}{dr^2}\delta G_{tr} +B\frac{d}{dr}\delta G_{tr}+C\delta
G_{tr}=0,
\end{equation}
where $A,B$ and $C$ were given by (20) in~\cite{Hirayama:2001bi,Myung:2013uka}.
See Fig. 2 that is  obtained   by solving Eq.(\ref{secondG-eq})  numerically.  We note that the small black holes of  horizon radii $r_+=1,2,4$ correspond to
$r_0=1.01,2.08,4.64$ with $\ell=10$, respectively.  $M$ ends up at $M^c\equiv {\cal O}(1)/r_0|_{r_+=1,2,4}=0.84,0.40,0.18$ whose horizontal lines appear in Fig. 1. From the observation of Fig. 2
with ${\cal O}(1)\simeq 0.85$, we find unstable modes for given $r_+=1,2,4$ as
\begin{equation} \label{unst-con}
0<M<M^c=\frac{{\cal O}(1)}{r_0}. \end{equation}
As the horizon size $r_+$ increases, the instability region  becomes narrow and narrow.
We call this instability as the Gregory-Laflamme instability~\cite{Gregory:1993vy,Hirayama:2001bi} because the four-dimensional linearized equation for $h_{\mu\nu}$ around the five-dimensional  black string background leads to
(\ref{EWL-eq}) when replacing $h_{\mu\nu}$ by $\delta G_{\mu\nu}$.
We check that for $r_+=6>r_*=5.7$, the maximum value of $\Omega$ is less than $10^{-4}$, which implies that there is no unstable modes  for large black hole with $r_+>r_*$.
The case of $M=0$ yields the critical gravity avoiding massive spin-2 and spin-0 modes when choosing the transverse-traceless gauge for $h_{\mu\nu}$~\cite{Liu:2011kf}.
At the critical point, the massless spin-2 modes have zero energy whereas  the massive spin-2 modes are
replaced by the log modes. The presence of log modes implies another instability   of the SAdS black hole in critical gravity.

For given $r_+=1,2,3$, from Figs. 1 and 2,  the stable
condition of the SAdS black hole in Einstein-Weyl gravity is given by
\begin{equation}
M>M^c. \label{sads-lbh}
\end{equation}
It show clearly that  the Gregory-Laflamme
instability of small black holes in the Einstein-Weyl gravity is due to
the massiveness of $M \in (0,M^c)$, but not a feature of fourth-order gravity giving
ghost states.
Taking into account the number of degrees of freedom (DOF), it is
helpful to show that  the SAdS black hole is physically stable in
the Einstein gravity~\cite{Cardoso:2001bb,Ishibashi:2003ap},
whereas the small SAdS black hole is unstable in the Einstein-Weyl
gravity.   The number of DOF of the metric perturbation is two in the
Einstein gravity, while the number of DOF of massive spin-2 $\delta G_{\mu\nu}$  is five in the
Einstein-Weyl gravity. The $s(l=0)$-mode analysis of the massive spin-2 with five DOF shows the Gregory-Laflamme instability. The $s$-mode
analysis is useful for handling the massive spin-2 mode in the Einstein-Weyl
gravity, but is not suitable for  the massless spin-2 mode  in the Einstein gravity.

 \subsection{SAdS black hole in Einstein-Ricci cubic  gravity}

First of all, we consider the  $\alpha=-2\beta$  Einstein-Ricci cubic gravity.
We wish to solve Eq.(\ref{LE-3}) for the traceless Ricci tensor $\delta \hat{R}_{\mu\nu}$.
 Actually, it is observed  that  Eq.(\ref{LE-3}) becomes Eq.(\ref{EWL-eq}) when substituting $\delta \hat{R}_{\mu\nu}$ and $\mu_2^2$ by  $\delta G_{\mu\nu}$ and $M^2$.
However,  we note that $\delta G_{\mu\nu}$ is a transverse-traceless tensor in Einstein-Weyl gravity, while  $\delta \hat{R}_{\mu\nu}$ is a traceless tensor in the $\alpha=-2\beta$  Einstein-Ricci cubic gravity.
 Hence, the DOF of  $\delta G_{\mu\nu}$ are  five, whereas the DOF of  $\delta \hat{R}_{\mu\nu}$ is nine. The non-transversality for $\delta \hat{R}_{\mu\nu}$  does not make a further progress on the $s$-mode analysis of stability.

Now, let us consider the $\alpha=-3\beta$ Einstein-Ricci cubic gravity. Its linearized equation is given by Eq.(\ref{LE-5}) with the mass squared (\ref{MLE-1}).
Eq.(\ref{LE-5}) becomes Eq.(\ref{EWL-eq}) exactly  when substituting $\delta \hat{R}_{\mu\nu}$ and $M_2^2$ by  $\delta G_{\mu\nu}$ and $M^2$.
Importantly, the traceless Ricci tensor $\delta\hat{ R}_{\mu\nu}$ satisfies the transverse condition (\ref{MLE-3}). Hence, its DOF is determined to be five as for $\delta G_{\mu\nu}$ in Einstein-Weyl gravity.
 At this stage, the even-parity metric perturbation could be
chosen for a single $s$-mode analysis in the $\alpha=-3\beta$ Einstein-Ricci cubic gravity and
whose form is given by
\begin{eqnarray}
\delta \hat{R}^e_{\mu\nu}=e^{\Omega t} \left(
\begin{array}{cccc}
\delta \hat{R}_{tt}(r) & \delta \hat{R}_{tr}(r) & 0 & 0 \cr \delta \hat{R}_{tr}(r) &\delta  \hat{R}_{rr}(r) & 0 & 0 \cr
0 & 0 & \delta \hat{R}_{\theta\theta}(r) & 0 \cr 0 & 0 & 0 & \sin^2\theta \delta \hat{R}_{\theta\theta}(r)
\end{array}
\right). \label{evenp}
\end{eqnarray}
 \begin{figure*}[t!]
   \centering
   \includegraphics{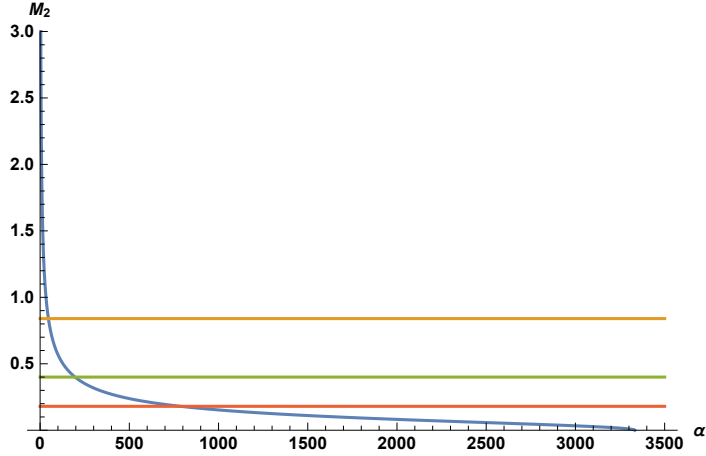}
\caption{Plots of mass $M_2$ for a massive spin-2 as a function of $\alpha$  with  $l=10$ and $\kappa=1$ in the $\alpha=-3\beta$  Einstein-Ricci cubic  gravity. Three horizontal lines correspond to $M_2^c={\cal O}(1)/r_0|_{r_+=1,2,4}=0.84,0.40,0.18$ where the critical gravity appears.
The mass $M_2$ is zero  at $\alpha=\alpha^*=3333$.  }
\end{figure*}
Once one starts with 4 DOF, they are  related to each other
when using  the Bianchi identity of $\bar{\nabla}^\mu\delta \hat{R}_{\mu\nu}=0$ together  with $\delta \hat{R}=0$. Thus, we  derive one
decoupled second-order equation for  $\delta \hat{R}_{tr}$,
\begin{equation} \label{secondhR-eq} A(r;r_0,\ell,\Omega^2,M_2^2)
\frac{d^2}{dr^2}\delta \hat{R}_{tr} +B\frac{d}{dr}\delta \hat{R}_{tr}+C\delta
\hat{R}_{tr}=0,
\end{equation}
where $A,B$ and $C$ were given by (20) in~\cite{Myung:2013uka}.
A physical mode of $\delta \hat{R}_{tr}$ grows exponentially in time as $e^{\Omega t}$ with $\Omega>0$, spatially vanishes at the AdS infinity, and regular at the future horizon~\cite{Hirayama:2001bi}.
Here $M_2$ is the mass of massive spin-2 mode  given by
\begin{equation} M_2=\sqrt{\frac{\ell^2}{3\alpha}-\frac{1}{\ell^2}}
\end{equation}
with $\kappa=1$.
We depict $M_2$ as a function of $\alpha$ in Fig. 3. The  massive spin-2 mass $M_2$ is zero at $\alpha=\alpha^*=\ell^4/3=3333
$, where the critical gravity appears.
We require $M^2_2>0$ to avoid the tachyonic instability of $\delta \hat{ R}_{\mu\nu}$ propagating on the SAdS black hole background~\cite{Liu:2011kf}. The tachyon-free condition implies the allowed range for $\alpha$ as
\begin{equation}
0<\alpha<\alpha^*.
\end{equation}
If $M^2_2<0(\alpha>\alpha^*)$, we do not need to perform a further analysis for the stability because it indicates  the tachyonic instability.

It is emphasized that  $M^2_2>0$ is not a necessary and sufficient condition to obtain a stable SAdS black hole.
We need to follow  Gregory-Laflamme scheme to distinguish between stable and unstable black holes by solving Eq.(\ref{secondhR-eq}) numerically.
Observing  Fig. 2 when replacing $M$ by $ M_2$, we note that  $M_2$ ends up at $M_2^c\equiv {\cal O}(1)/r_0|_{r_+=1,2,4}=0.84,0.40,0.18$ with ${\cal O}(1)=0.85$ whose horizontal lines appear in Fig. 3.
Here, we find unstable modes for small black holes with $r_+=1,2,4$ as
\begin{equation} \label{unst-con2}
0<M_2<M_2^c=\frac{{\cal O}(1)}{r_0}. \end{equation}
As the horizon size $r_+$ increases, the instability region starting from the origin  becomes narrow and narrow.
For given $r_+=1,2,4$, three horizontal lines $M_2^c$  which are ending points split unstable ($M_2<M_2^c$) and stable ($M_2>M_2^c$) black holes.

 It indicates  that  the Gregory-Laflamme
instability of small black holes in the $\alpha=-3\beta$ Einstein-Ricci cubic gravity is due to
the massiveness ($0<M_2<M_2^c$) of massive spin-2 mode, but not a feature of higher-order gravity giving
a ghost.  This ghost  may appear only when expressing the linearized equation (\ref{LE-5}) in terms of the  metric perturbation $h_{\mu\nu}$, giving seven DOF.
Here the massive ghost (an unhealthy  massive spin-2 mode) does not appear because we used the linearized Ricci tensor $\delta \hat{R}_{\mu\nu}$ to represent a healthy massive spin-2 mode. However, this does not mean that our perturbation analysis misses the ghost instability. The ghost is present in the spectrum of the theory because the theory provides a fourth-order linearized equation when expressing in terms of $h_{\mu\nu}$. The matter is how to represent a massive spin-2 mode propagating on the SAdS black hole spacetimes. Expressing  the linearized equation in term of the linearized Ricci tensor instead of the metric perturbation, it becomes a second-order linearized equation. This is one tip to handle a fourth-order linearized gravity.
Considering an auxiliary field formulation for decreasing a fourth-order gravity to a second-order theory of gravity~\cite{Mauro:2015waa},
one found that the linearized equation for Ricci tensor $\delta R_{\mu\nu}$ is transformed exactly into that for  auxiliary field $\psi_{\mu\nu}$.
Using an auxiliary field enables one to distinguish the perturbation related to a massive ghost. Furthermore, we wish to point out that the linearized Ricci tensor could represent a massive spin-2 mode with five DOF~\cite{Eardley:1974nw,Moon:2014qma}.
We stress here that the Gregory-Laflamme instability based on $\delta R_{\mu\nu}$ has nothing to do with the ghost instability and it reflects a feature of massive gravity described by the linearized equation (\ref{LE-5}).

\section{Wald entropy and black hole thermodynamics}
It is known that the correlated  stability conjecture proposed by Gubser-Mitra~\cite{Gubser:2000ec}  does not hold for the SAdS black hole found
 in Einstein gravity, but it holds for the SAdS black hole found in Einstein-Weyl gravity~\cite{Myung:2013uka}. In order to confirm the  classical instability found in the previous section,
 one has to explore the thermodynamic property of the SAdS black hole obtained from  the Einstein-Ricci cubic gravity.

\subsection{Einstein-Weyl gravity}
We start with the Einstein-Weyl gravity because the thermodynamic quantities  of the SAdS black hole was completely computed  by employing  the Abbot-Deser-Tekin method~\cite{Lu:2011zk}.
It was well-known that  the Wald entropy of Einstein-Weyl gravity (\ref{EWG}) is given by
\begin{eqnarray}  \label{entropy2}
S_{\rm W}^{\rm EW}= \frac{A}{4}\Big[1+2b\Lambda\Big]= \Big[1-\frac{6b}{\ell^2}\Big]S_{\rm BH}  \label{entropy3}
\end{eqnarray}
with $S_{\rm BH}=\pi r_+^2$ the Bekenstein-Hawking entropy of the SAdS black hole in the Einstein gravity.
The other thermodynamic quantities of mass, heat capacity, and  free energy  are given by~\cite{Lu:2011zk,Myung:2013uka}
\begin{equation}
M^{\rm EW}=\Big[1-\frac{6b}{\ell^2}\Big]M_{\rm SAdS},~C^{\rm EW}=\Big[1-\frac{6b}{\ell^2}\Big]C_{\rm SAdS},~F^{\rm EW}=\Big[1-\frac{6b}{\ell^2}\Big]F_{\rm SAdS}, \label{ther-q1}
\end{equation}
where the thermodynamic quantities including the Hawking temperature for a SAdS black hole in Einstein gravity take the forms
\begin{eqnarray}
M_{\rm SAdS}&=&\frac{1}{2}\Big[r_++\frac{r_+^3}{\ell^2}\Big],~~C_{\rm SAdS}=2\pi r_+^2\Big[\frac{3r_+^2+\ell^2}{3r_+^2-\ell^2}\Big], \nonumber \\
F_{\rm SAdS}&=&\frac{1}{4}\Big[r_+-\frac{r_+^3}{\ell^2}\Big],~~T_{\rm H}=\frac{1}{4\pi}\Big[\frac{1}{r_+}+\frac{3r_+}{\ell^2}\Big].\label{ther-q2}
\end{eqnarray}
We check that the first-law of thermodynamics is satisfied in Einstein-Weyl gravity as
\begin{equation}
dM^{\rm EW}=T_{\rm H}dS_{\rm W}^{\rm EW},
\end{equation}
as the first-law is indeed  satisfied in Einstein gravity
\begin{equation}
dM_{\rm SAdS}=T_{\rm H}dS_{\rm BH},
\end{equation}
where $``d"$ denotes the differentiation with respect to the horizon radius $r_+$ only.
\begin{figure*}[t!]
  \centering
 \includegraphics{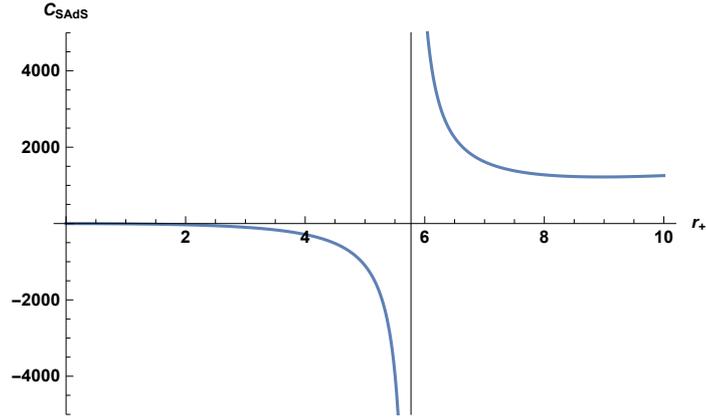}
\caption{Plot of heat capacity $C_{\rm SAdS}$ with $l=10$ in Einstein gravity.  The heat capacity blows up at $r_+=r_*=\ell/\sqrt{3}=5.7$. The thermodynamic stability is based on the sign of  heat capacity.  The small (unstable) black hole with $r_+<r_*$ is defined by the negative heat capacity,
whereas the large (stable) black hole with $r_+>r_*$ is defined by the positive heat capacity. This picture persists to  $C^{\rm EW}$ with $b<\ell^2/6$ and $C^{\alpha=-3\beta}$ with $\alpha<\ell^4/3$.}
\end{figure*}

We briefly sketch the thermodynamic stability of the SAdS black hole found in  Einstein-Weyl gravity~\cite{Myung:2013uka}.
First we consider the case of $b<\ell^2/6(M^2>0)$ which is dominantly described by the Einstein-Hilbert term.
Since the heat capacity $C_{\rm SAdS}$ blows up at $r_+=r_*=\ell/\sqrt{3}=5.7$ [see Fig. 4], we divide the black hole into the small black hole with $r_+<r_*$
and the large black hole with  $r_+>r_*$. We know that the small black hole is thermodynamically unstable because the heat capacity $C^{\rm EW}<0$, while the large black hole is thermodynamically stable  because  $C^{\rm EW}>0$.
 For the other case of $b>\ell^2/6(M^2<0)$ which is dominantly described by the Weyl term, the situation reverses. The small black hole  is thermodynamically stable because  $C^{\rm EW}>0$, while the large black hole is thermodynamically unstable  because  $C^{\rm EW}<0$. In this case,   the mass squared  of massive spin-2 is negative, which implies the tachyonic instability. So, this case should be excluded from the consideration.
We would like to mention that there is no connection between classical stability and thermodynamic instability for small SAdS black hole in Einstein gravity.
However, let us see how things are improved in Einstein-Weyl gravity. For a small black hole with $r_+<r_*$ and $b/\ell^2/6$, the heat capacity is negative which means that it is thermodynamically unstable.
On the other hand, we observe  from (\ref{unst-con}) that a small black hole with $r_+<r_*$ is unstable against the $s$-mode massive spin-2 perturbation. The Gregory-Laflamme instability condition picks up the small SAdS black hole which is thermodynamically unstable in Einstein-Weyl gravity.
 This implies that the CSC~\cite{Gubser:2000ec} holds for the SAdS black hole found in Einstein-Weyl gravity.

\subsection{Einstein-Ricci cubic gravity}
First of all,  we wish to compute the Wald entropy of the SAdS black hole in Einstein-Ricci cubic gravity.
 The Wald entropy is defined by the following integral performed on 2-dimensional spacelike bifurcation surface $\Sigma$~\cite{Wald:1993nt,Iyer:1994ys,Jacobson:1993vj}:
 \begin{equation} \label{w-ent}
 S_{\rm W}=-\frac{1}{8} \oint \Big(\frac{\delta {\cal L}_{\rm ER}}{\delta R_{\mu\nu\rho\sigma}}\Big)^{(0)} \epsilon_{\mu\nu}\epsilon_{\rho\sigma}dV^2_2=-\frac{1}{8} \oint \bar{P}^{\mu\nu\rho\sigma}\epsilon_{\mu\nu}\epsilon_{\rho\sigma}dV^2_2,
 \end{equation}
 where $dV^2_2=r^2\sin \theta d\theta d\phi$ is the volume element on $\Sigma$ and $\epsilon_{\mu\nu}$ is the binormal vector to $\Sigma$ normalized as $ \epsilon_{\mu\nu} \epsilon^{\mu\nu}=-2$.
 We note that  the superscript (0) denotes that the functional derivative with respect to $R_{\mu\nu\rho\sigma}$ is evaluated  on-shell.
 The background (on-shell) $P$-tensor $\bar{P}^{\mu\nu\rho\sigma}$ is given by (\ref{p-tensor}).
Now, the Wald entropy takes the form
\begin{eqnarray}  \label{entropy1}
S_{\rm W}&=&\frac{1}{4}\oint\Big[1+(\alpha+4\beta)\Lambda^2\Big]r^2_+\sin\theta d\theta d\phi \nonumber \\
&=& \frac{A}{4}\Big[1+(\alpha+4\beta)\Lambda^2\Big] \nonumber \\
\end{eqnarray}
 with $ A=4\pi r^2_+$ and $\kappa=1/G=1$.
In the case of $\alpha=-3\beta$ Einstein-Ricci cubic gravity, the Wald entropy takes the form
\begin{eqnarray}
S_{\rm W}^{\alpha=-3\beta}= \frac{A}{4}\Big[1-\frac{\alpha}{3}\Lambda^2\Big]=  \Big[1-\frac{3\alpha}{\ell^4}\Big]S_{\rm BH}. \label{entropy2}
\end{eqnarray}

Up to now, we  know the Wald entropy $S^{\alpha=-3\beta}_{\rm W}$ and  the Hawking temperature
\begin{equation} \label{H-temp}
T_{\rm H}=\frac{1}{4\pi}\Big[\frac{1}{r_+}+\frac{3r_+}{\ell^2}\Big].
\end{equation}
However,  we do not know the other thermodynamic  quantities of the SAdS black hole in Einstein-Ricci cubic gravity.
We propose that the  first-law should be satisfied in $\alpha=-3\beta$ Einstein-Ricci cubic gravity as
\begin{equation} \label{1st-law}
dM^{\rm \alpha=-3\beta}=T_{\rm H}dS_{\rm W}^{\rm \alpha=-3\beta}.
\end{equation}
Using (\ref{1st-law}) together with (\ref{entropy2}) and (\ref{H-temp}), we derive the mass of the SAdS black hole
\begin{equation}\label{mass-bh}
  M^{\alpha=-3\beta }(r_+)=\int^{r_{+}}_0 dr_+'T_{\rm H}(r_+')dS_{\rm W}^{\alpha=-3\beta}(r_+')=\Big[1-\frac{3\alpha}{\ell^4}\Big]M_{\rm SAdS}(r_+).
\end{equation}
Other  thermodynamic quantities of heat capacity and free energy are computed to be
\begin{eqnarray} \label{entropy3}
C^{ \alpha=-3\beta }(r_+)&=&\Big(\frac{dM^{\alpha=-3\beta }}{dT_{\rm H}}\Big)=\Big[1-\frac{3\alpha}{\ell^4}\Big]C_{\rm SAdS}(r_+),\\
F^{ \alpha=-3\beta }(r_+)&=& M^{\alpha=-3\beta }-T_{\rm H}S_{\rm W}^{\alpha=-3\beta}=\Big[1-\frac{3\alpha}{\ell^4}\Big]F_{\rm SAdS}(r_+). \label{freeE}
\end{eqnarray}
Now we are in a position  to  mention the thermodynamic stability of the SAdS black hole in $\alpha=-3\beta$ Einstein-Ricci cubic gravity.
First, we consider the case of $\alpha<\ell^4/3<1(M_2^2>0)$ which is dominantly described by the Einstein-Hilbert  term.
Since the heat capacity $C^{\alpha=-3\beta}$ blows up at $r_+=r_*=\ell/\sqrt{3}=0.57\ell$ [see Fig. 4], we divide still the black hole into the small black hole with $r_+<r_*$
and the large black hole with  $r_+>r_*$. Then, it suggests  that the small black hole is thermodynamically unstable because  $C^{\alpha=-3\beta}<0$, while the large black hole is thermodynamically stable  because  $C^{\alpha=-3\beta}>0$. For the other case of $\alpha>\ell^4/3(M_2^2<0)$ which is dominantly described by the Ricci cubic terms, the thermodynamic stability reverses. The small black hole  is thermodynamically stable because
$C^{\alpha=-3\beta}>0$, while the large black hole is thermodynamically unstable  because  $C^{\alpha=-3\beta}<0$. However,  we note that the mass squared $M_2^2$ of massive spin-2 mode is negative.
This corresponds to the tachyonic instability and thus,  this case is unacceptable.

One finds from (\ref{unst-con2}) that for $M^2_2>0$, a small (large) black hole with $r_+<r_*(r_+>r_*)$ is unstable (stable)  against the $s$-mode massive spin-2 perturbation $\delta \hat{R}_{\mu\nu}$. The Gregory-Laflamme instability condition picks up the small SAdS black hole which is thermodynamically unstable in $\alpha=-3\beta$ Einstein-Ricci cubic gravity.
This indicates that the CSC proposed by Gubser-Mitra~\cite{Gubser:2000ec} holds for the SAdS black hole found
in $\alpha=-3\beta$ Einstein-Ricci cubic  gravity too. However, the other case of $M^2_2<0$ corresponds to the tachyonic instability and its thermodynamic stability contradicts to the conventional one. Hence, the CSC does not hold for $M^2_2<0$.

\section{Discussions}
We would like to mention  the Einstein-Ricci cubic gravity according to the relation between $\alpha=4e_2+3e_3$ and $\beta=2(6e_1+2e_2)$. \\
i) $\alpha =\beta=0$ case.
This case describes  a quasi-topological Ricci cubic gravity  because its linearized equation around the SAdS black hole reduces to that of  Einstein gravity with cosmological constant. There is no counterpart in fourth-order gravity. The absence of massive spin-0 and spin-2  modes implies that the fourth-order terms give no contribution to this linearized theory as if they were purely topological, implying that  this linearized theory is ghost-free. The SAdS black hole is stable regardless of the horizon radius $r_+$ in  $\alpha =\beta=0$ Einstein-Ricci cubic gravity because its linearized equation takes the form of $\delta G_{\mu\nu}(h)=0$ without any mass terms. However, the small black hole with $r_+<r_*$ is thermodynamically unstable while the large black hole with $r_+>r_*$ is thermodynamically unstable.  The CSC does not hold for the $\alpha=\beta=0$ Einstein-Ricci cubic gravity which is a ghost-free theory.\\
ii) $\alpha =-2\beta$ case.
This case allows a decoupling of  massive spin-0 mode from massive spin-0 mode.
This corresponds to the fourth-order gravity. Here $\delta \hat{R}_{\mu\nu}$ is  only a traceless tensor whose DOF is nine. We note that the absence of transversality condition $\bar{\nabla}^\mu \delta \hat{R}_{\mu\nu}=0$ makes the stability analysis of SAdS black hole difficult. Its Wald entropy is given by $S^{\alpha=-2\beta}_{\rm W}=S_{\rm BH}[1-\alpha \Lambda^2]$. All thermodynamic quantities are similar to those for $\alpha=-3\beta$ Einstein-Ricci cubic gravity except replacing $[1-\alpha \Lambda^2/3]$ by $[1-\alpha \Lambda^2]$.   For $\alpha<\ell^4/9$, the small black hole with $r_+<r_*$ is thermodynamically unstable while the large black hole with $r_+>r_*$ is thermodynamically unstable.  In this case, we may not discuss the CSC because we could not explore the stability issue of the SAdS black hole explicitly in $\alpha =-2\beta$ Einstein-Ricci cubic gravity.\\
iii) $\alpha =-3\beta$ case.
This is related  to the Einstein-Weyl  gravity. We show by adopting the Gregory-Laflamme scheme that the small SAdS black hole with $r_+<r_*$ is classically unstable, while the large SAdS black hole with $r_+>r_*$ is classically stable. It indicates clearly that the Gregory-Laflamme instability arises from the massiveness ($0<M_2<M_2^c$) of massive spin-2 mode, but not from a feature of Einstein-Ricci cubic (fourth-order) gravity giving ghost states. On the other hand, its thermodynamic quantities are  computed   by  making use of the Wald entropy and the first-law of thermodynamics.  For the case of $M^2_2>0$ which is dominantly described by the Einstein-Hilbert term, one finds that the small black hole with $r_+<r_*$ is thermodynamically unstable, whereas the large  black hole with $r_+>r_*$ is thermodynamically stable. This shows that the CSC proposed by Gubser-Mitra~\cite{Gubser:2000ec} works for the SAdS black hole obtained  from the $\alpha =-3\beta$ Einstein-Ricci cubic gravity.  The other case of $M^2_2<0$ corresponds to the tachyonic instability and its thermodynamic stability is unacceptable. Hence, the CSC does not hold for $M^2_2<0$.\\
iv) $\alpha=-3\beta$  case with $M^2_2=0(\alpha=\alpha^*)$.
In this case, considering the transverse and traceless gauge of $\bar{\nabla}^\mu h_{\mu\nu}=0$ and $h=0$, the linearized Einstein tensor takes the from of $\delta G_{\mu\nu}=-(\bar{\Delta}_{\rm L}-2\Lambda)h_{\mu\nu}/2$. Then, its linearized equation is given by $(\bar{\Delta}_{\rm L}-2\Lambda)^2h_{\mu\nu}=0$ that corresponds to the critical gravity.  It turned out be an unstable theory, even though massive spin-2 and massive spin-0 modes are decoupled from the theory~\cite{Liu:2011kf}. On the other hand, all thermodynamic quantities disappear except the Hawking temperature. Therefore, we could not discuss the CSC for this case.

Consequently, we have obtained the SAdS black hole solution (\ref{sch}) from the Einstein-Ricci cubic gravity (\ref{Action}).
 A physical black hole could be found by performing the analysis of stability which means a perturbation analysis based on physical field around the black hole because the black hole  is a solution in the curved spacetimes~\cite{chan,Kwon:1986dw}.
In this work, we have shown that  the $\alpha=-3\beta$ Einstein-Ricci cubic gravity passes  the stability analysis of SAdS black hole.
The $\alpha=\beta=0$ case provides a quasi-topological Ricci cubic gravity whose linearized theory is just the linearized Einstein gravity,
 implying that the black hole solution (\ref{sch}) becomes a physical black hole in $\alpha=\beta=0$  Einstein-Ricci cubic gravity.
However, we could not establish the (in) stability for  other relations between  $\alpha$ and $\beta$.
In addition, it is worth noting that the $\alpha=-3\beta$ Einstein-Ricci cubic gravity has a similar property to the Einstein-Weyl gravity because any higher-order gravity could be  mapped  into
the quadratic gravity at the linearized level.

Finally, if one considers the Riemann cubic gravity (Einsteinian cubic gravity), covariant linearized gravity is possible only in maximally symmetric vacua.  Also,  black hole solutions are  numerical or approximate solutions in the Einsteinian cubic gravity and an obstacle to studying these black hole is the lack of an analytic solution~\cite{Hennigar:2018hza}.
 \vspace{1cm}

{\bf Acknowledgments}
 \vspace{1cm}

This work was supported by the National Research Foundation of Korea (NRF) grant funded by the Korea government (MOE)
 (No. NRF-2017R1A2B4002057).


\vspace{1cm}

\begin{thebibliography}{99}

%\cite{Stelle:1976gc}
\bibitem{Stelle:1976gc}
  K.~S.~Stelle,
  %``Renormalization of Higher Derivative Quantum Gravity,''
  Phys.\ Rev.\ D {\bf 16}, 953 (1977).
  doi:10.1103/PhysRevD.16.953
  %%CITATION = doi:10.1103/PhysRevD.16.953;%%
  %1469 citations counted in INSPIRE as of 11 Jan 2018

%\cite{Modesto:2011kw}
\bibitem{Modesto:2011kw}
  L.~Modesto,
  %``Super-renormalizable Quantum Gravity,''
  Phys.\ Rev.\ D {\bf 86}, 044005 (2012)
  doi:10.1103/PhysRevD.86.044005
  [arXiv:1107.2403 [hep-th]].
  %%CITATION = doi:10.1103/PhysRevD.86.044005;%%
  %221 citations counted in INSPIRE as of 11 Jan 2018

%\cite{Biswas:2011ar}
\bibitem{Biswas:2011ar}
  T.~Biswas, E.~Gerwick, T.~Koivisto and A.~Mazumdar,
  %``Towards singularity and ghost free theories of gravity,''
  Phys.\ Rev.\ Lett.\  {\bf 108}, 031101 (2012)
  doi:10.1103/PhysRevLett.108.031101
  [arXiv:1110.5249 [gr-qc]].
  %%CITATION = doi:10.1103/PhysRevLett.108.031101;%%
  %281 citations counted in INSPIRE as of 16 Nov 2017

%\cite{Lovelock:1971yv}
\bibitem{Lovelock:1971yv}
  D.~Lovelock,
  %``The Einstein tensor and its generalizations,''
  J.\ Math.\ Phys.\  {\bf 12}, 498 (1971).
  doi:10.1063/1.1665613
  %%CITATION = doi:10.1063/1.1665613;%%
  %1322 citations counted in INSPIRE as of 11 Jan 2018


%\cite{Myers:2010ru}
\bibitem{Myers:2010ru}
  R.~C.~Myers and B.~Robinson,
  %``Black Holes in Quasi-topological Gravity,''
  JHEP {\bf 1008}, 067 (2010)
  doi:10.1007/JHEP08(2010)067
  [arXiv:1003.5357 [gr-qc]].
  %%CITATION = doi:10.1007/JHEP08(2010)067;%%
  %121 citations counted in INSPIRE as of 11 Jan 2018

%\cite{Cisterna:2017umf}
\bibitem{Cisterna:2017umf}
  A.~Cisterna, L.~Guajardo, M.~Hassaine and J.~Oliva,
  %``Quintic quasi-topological gravity,''
  JHEP {\bf 1704}, 066 (2017)
  doi:10.1007/JHEP04(2017)066
  [arXiv:1702.04676 [hep-th]].
  %%CITATION = doi:10.1007/JHEP04(2017)066;%%
  %11 citations counted in INSPIRE as of 11 Jan 2018


%\cite{Oliva:2010eb}
\bibitem{Oliva:2010eb}
  J.~Oliva and S.~Ray,
  %``A new cubic theory of gravity in five dimensions: Black hole, Birkhoff's theorem and C-function,''
  Class.\ Quant.\ Grav.\  {\bf 27}, 225002 (2010)
  doi:10.1088/0264-9381/27/22/225002
  [arXiv:1003.4773 [gr-qc]].
  %%CITATION = doi:10.1088/0264-9381/27/22/225002;%%
  %85 citations counted in INSPIRE as of 11 Jan 2018

%\cite{Bueno:2016xff}
\bibitem{Bueno:2016xff}
  P.~Bueno and P.~A.~Cano,
  %``Einsteinian cubic gravity,''
  Phys.\ Rev.\ D {\bf 94}, no. 10, 104005 (2016)
  doi:10.1103/PhysRevD.94.104005
  [arXiv:1607.06463 [hep-th]].
  %%CITATION = doi:10.1103/PhysRevD.94.104005;%%
  %22 citations counted in INSPIRE as of 11 Jan 2018

%\cite{Hennigar:2017ego}
\bibitem{Hennigar:2017ego}
  R.~A.~Hennigar, D.~Kubiznak and R.~B.~Mann,
  %``Generalized quasitopological gravity,''
  Phys.\ Rev.\ D {\bf 95}, no. 10, 104042 (2017)
  doi:10.1103/PhysRevD.95.104042
  [arXiv:1703.01631 [hep-th]].
  %%CITATION = doi:10.1103/PhysRevD.95.104042;%%
  %11 citations counted in INSPIRE as of 11 Jan 2018

%\cite{Ahmed:2017jod}
\bibitem{Ahmed:2017jod}
  J.~Ahmed, R.~A.~Hennigar, R.~B.~Mann and M.~Mir,
  %``Quintessential Quartic Quasi-topological Quartet,''
  JHEP {\bf 1705}, 134 (2017)
  doi:10.1007/JHEP05(2017)134
  [arXiv:1703.11007 [hep-th]].
  %%CITATION = doi:10.1007/JHEP05(2017)134;%%
  %7 citations counted in INSPIRE as of 11 Jan 2018

%\cite{Bueno:2016lrh}
\bibitem{Bueno:2016lrh}
  P.~Bueno and P.~A.~Cano,
  %``Four-dimensional black holes in Einsteinian cubic gravity,''
  Phys.\ Rev.\ D {\bf 94}, no. 12, 124051 (2016)
  doi:10.1103/PhysRevD.94.124051
  [arXiv:1610.08019 [hep-th]].
  %%CITATION = doi:10.1103/PhysRevD.94.124051;%%
  %14 citations counted in INSPIRE as of 12 Jan 2018

%\cite{Feng:2017tev}
\bibitem{Feng:2017tev}
  X.~H.~Feng, H.~Huang, Z.~F.~Mai and H.~Lu,
  %``Bounce Universe and Black Holes from Critical Einsteinian Cubic Gravity,''
  Phys.\ Rev.\ D {\bf 96}, no. 10, 104034 (2017)
  doi:10.1103/PhysRevD.96.104034
  [arXiv:1707.06308 [hep-th]].
  %%CITATION = doi:10.1103/PhysRevD.96.104034;%%
  %3 citations counted in INSPIRE as of 02 Apr 2018



%\cite{Li:2017ncu}
\bibitem{Li:2017ncu}
  Y.~Z.~Li, H.~S.~Liu and H.~Lu,
  %``Quasi-Topological Ricci Polynomial Gravities,''
  arXiv:1708.07198 [hep-th].
  %%CITATION = ARXIV:1708.07198;%%
  %8 citations counted in INSPIRE as of 10 Jan 2018



%\cite{Bueno:2016ypa}
\bibitem{Bueno:2016ypa}
  P.~Bueno, P.~A.~Cano, V.~S.~Min and M.~R.~Visser,
  %``Aspects of general higher-order gravities,''
  Phys.\ Rev.\ D {\bf 95}, no. 4, 044010 (2017)
  doi:10.1103/PhysRevD.95.044010
  [arXiv:1610.08519 [hep-th]].
  %%CITATION = doi:10.1103/PhysRevD.95.044010;%%
  %22 citations counted in INSPIRE as of 12 Jan 2018

%\cite{Bueno:2017sui}
\bibitem{Bueno:2017sui}
  P.~Bueno and P.~A.~Cano,
  %``On black holes in higher-derivative gravities,''
  Class.\ Quant.\ Grav.\  {\bf 34}, no. 17, 175008 (2017)
  doi:10.1088/1361-6382/aa8056
  [arXiv:1703.04625 [hep-th]].
  %%CITATION = doi:10.1088/1361-6382/aa8056;%%
  %10 citations counted in INSPIRE as of 02 Apr 2018

%\cite{Myung:2013bow}
\bibitem{Myung:2013bow}
  Y.~S.~Myung,
  %``Instability of Schwarzschild-AdS black hole in Einstein-Weyl gravity,''
  Phys.\ Lett.\ B {\bf 728}, 422 (2014)
  doi:10.1016/j.physletb.2013.12.019
  [arXiv:1308.1455 [gr-qc]].
  %%CITATION = doi:10.1016/j.physletb.2013.12.019;%%
  %5 citations counted in INSPIRE as of 10 Jan 20

%\cite{Regge:1957td}
\bibitem{Regge:1957td}
  T.~Regge and J.~A.~Wheeler,
  %``Stability of a Schwarzschild singularity,''
  Phys.\ Rev.\  {\bf 108}, 1063 (1957).
  doi:10.1103/PhysRev.108.1063
  %%CITATION = doi:10.1103/PhysRev.108.1063;%%
  %1205 citations counted in INSPIRE as of 10 Jan 2018

%\cite{Zerilli:1970se}
\bibitem{Zerilli:1970se}
  F.~J.~Zerilli,
  %``Effective potential for even parity Regge-Wheeler gravitational perturbation equations,''
  Phys.\ Rev.\ Lett.\  {\bf 24}, 737 (1970).
  doi:10.1103/PhysRevLett.24.737
  %%CITATION = doi:10.1103/PhysRevLett.24.737;%%
  %498 citations counted in INSPIRE as of 10 Jan 2018

%\cite{Vishveshwara:1970cc}
\bibitem{Vishveshwara:1970cc}
  C.~V.~Vishveshwara,
  %``Stability of the schwarzschild metric,''
  Phys.\ Rev.\ D {\bf 1}, 2870 (1970).
  doi:10.1103/PhysRevD.1.2870
  %%CITATION = doi:10.1103/PhysRevD.1.2870;%%
  %281 citations counted in INSPIRE as of 10 Jan 2018

\bibitem{chan} S. Chandrasekhar, in The Mathematical Theory of Black Holes
(Oxford University, New York, 1983).

%\cite{Cardoso:2001bb}
\bibitem{Cardoso:2001bb}
  V.~Cardoso and J.~P.~S.~Lemos,
  %``Quasinormal modes of Schwarzschild anti-de Sitter black holes: Electromagnetic and gravitational perturbations,''
  Phys.\ Rev.\ D {\bf 64}, 084017 (2001)
  doi:10.1103/PhysRevD.64.084017
  [gr-qc/0105103].
  %%CITATION = doi:10.1103/PhysRevD.64.084017;%%
  %240 citations counted in INSPIRE as of 10 Jan 2018

%\cite{Ishibashi:2003ap}
\bibitem{Ishibashi:2003ap}
  A.~Ishibashi and H.~Kodama,
  %``Stability of higher dimensional Schwarzschild black holes,''
  Prog.\ Theor.\ Phys.\  {\bf 110}, 901 (2003)
  doi:10.1143/PTP.110.901
  [hep-th/0305185].
  %%CITATION = doi:10.1143/PTP.110.901;%%
  %229 citations counted in INSPIRE as of 10 Jan 2018






%\cite{Whitt:1985ki}
\bibitem{Whitt:1985ki}
  B.~Whitt,
  %``The Stability of Schwarzschild Black Holes in Fourth Order Gravity,''
  Phys.\ Rev.\ D {\bf 32}, 379 (1985).
  doi:10.1103/PhysRevD.32.379
  %%CITATION = doi:10.1103/PhysRevD.32.379;%%
  %25 citations counted in INSPIRE as of 10 Jan 2018

%\cite{Mauro:2015waa}
\bibitem{Mauro:2015waa}
  S.~Mauro, R.~Balbinot, A.~Fabbri and I.~L.~Shapiro,
  %``Fourth derivative gravity in the auxiliary fields representation and application to the black hole stability,''
  Eur.\ Phys.\ J.\ Plus {\bf 130}, no. 7, 135 (2015)
  doi:10.1140/epjp/i2015-15135-0
  [arXiv:1504.06756 [gr-qc]].
  %%CITATION = doi:10.1140/epjp/i2015-15135-0;%%
  %4 citations counted in INSPIRE as of 12 Jan 2018


  %\cite{Stelle:2017bdu}
\bibitem{Stelle:2017bdu}
  K.~S.~Stelle,
  %``Abdus Salam and quadratic curvature gravity: Classical solutions,''
  Int.\ J.\ Mod.\ Phys.\ A {\bf 32}, no. 09, 1741012 (2017).
  doi:10.1142/S0217751X17410123
  %%CITATION = doi:10.1142/S0217751X17410123;%%
  %3 citations counted in INSPIRE as of 10 Jan 2018




  %\cite{Myung:2013doa}
\bibitem{Myung:2013doa}
  Y.~S.~Myung,
  %``Stability of Schwarzschild black holes in fourth-order gravity revisited,''
  Phys.\ Rev.\ D {\bf 88}, no. 2, 024039 (2013)
  doi:10.1103/PhysRevD.88.024039
  [arXiv:1306.3725 [gr-qc]].
  %%CITATION = doi:10.1103/PhysRevD.88.024039;%%
  %18 citations counted in INSPIRE as of 10 Jan 2018


%\cite{Gregory:1993vy}
\bibitem{Gregory:1993vy}
  R.~Gregory and R.~Laflamme,
  %``Black strings and p-branes are unstable,''
  Phys.\ Rev.\ Lett.\  {\bf 70}, 2837 (1993)
  doi:10.1103/PhysRevLett.70.2837
  [hep-th/9301052].
  %%CITATION = doi:10.1103/PhysRevLett.70.2837;%%
  %829 citations counted in INSPIRE as of 10 Jan 2018

%\cite{Gubser:2000ec}
\bibitem{Gubser:2000ec}
  S.~S.~Gubser and I.~Mitra,
  %``Instability of charged black holes in Anti-de Sitter space,''
  hep-th/0009126.
  %%CITATION = HEP-TH/0009126;%%
  %212 citations counted in INSPIRE as of 10 Jan 2018


%\cite{Harmark:2007md}
\bibitem{Harmark:2007md}
  T.~Harmark, V.~Niarchos and N.~A.~Obers,
  %``Instabilities of black strings and branes,''
  Class.\ Quant.\ Grav.\  {\bf 24}, R1 (2007)
  doi:10.1088/0264-9381/24/8/R01
  [hep-th/0701022].
  %%CITATION = doi:10.1088/0264-9381/24/8/R01;%%
  %156 citations counted in INSPIRE as of 11 Jan 2018


%\cite{deRham:2010kj}
\bibitem{deRham:2010kj}
  C.~de Rham, G.~Gabadadze and A.~J.~Tolley,
  %``Resummation of Massive Gravity,''
  Phys.\ Rev.\ Lett.\  {\bf 106}, 231101 (2011)
  doi:10.1103/PhysRevLett.106.231101
  [arXiv:1011.1232 [hep-th]].
  %%CITATION = doi:10.1103/PhysRevLett.106.231101;%%
  %978 citations counted in INSPIRE as of 22 Feb 2018

%\cite{Babichev:2013una}
\bibitem{Babichev:2013una}
  E.~Babichev and A.~Fabbri,
  %``Instability of black holes in massive gravity,''
  Class.\ Quant.\ Grav.\  {\bf 30}, 152001 (2013)
  doi:10.1088/0264-9381/30/15/152001
  [arXiv:1304.5992 [gr-qc]].
  %%CITATION = doi:10.1088/0264-9381/30/15/152001;%%
  %79 citations counted in INSPIRE as of 22 Feb 2018

%\cite{Brito:2013wya}
\bibitem{Brito:2013wya}
  R.~Brito, V.~Cardoso and P.~Pani,
  %``Massive spin-2 fields on black hole spacetimes: Instability of the Schwarzschild and Kerr solutions and bounds on the graviton mass,''
  Phys.\ Rev.\ D {\bf 88}, no. 2, 023514 (2013)
  doi:10.1103/PhysRevD.88.023514
  [arXiv:1304.6725 [gr-qc]].
  %%CITATION = doi:10.1103/PhysRevD.88.023514;%%
  %124 citations counted in INSPIRE as of 22 Feb 2018



%\cite{Myung:2013uka}
\bibitem{Myung:2013uka}
  Y.~S.~Myung and T.~Moon,
  %``Thermodynamic and classical instability of AdS black holes in fourth-order gravity,''
  JHEP {\bf 1404}, 058 (2014)
  doi:10.1007/JHEP04(2014)058
  [arXiv:1311.6985 [hep-th]].
  %%CITATION = doi:10.1007/JHEP04(2014)058;%%
  %6 citations counted in INSPIRE as of 10 Jan 2018





%\cite{Lu:2017kzi}
\bibitem{Lu:2017kzi}
  H.~Lu, A.~Perkins, C.~N.~Pope and K.~S.~Stelle,
  %``Lichnerowicz Modes and Black Hole Families in Ricci Quadratic Gravity,''
  Phys.\ Rev.\ D {\bf 96}, no. 4, 046006 (2017)
  doi:10.1103/PhysRevD.96.046006
  [arXiv:1704.05493 [hep-th]].
  %%CITATION = doi:10.1103/PhysRevD.96.046006;%%
  %5 citations counted in INSPIRE as of 10 Jan 2018



%\cite{Liu:2011kf}
\bibitem{Liu:2011kf}
  H.~Liu, H.~Lu and M.~Luo,
  %``On Black Hole Stability in Critical Gravities,''
  Int.\ J.\ Mod.\ Phys.\ D {\bf 21}, 1250020 (2012)
  doi:10.1142/S0218271812500204
  [arXiv:1104.2623 [hep-th]].
  %%CITATION = doi:10.1142/S0218271812500204;%%
  %22 citations counted in INSPIRE as of 10 Jan 2018




%\cite{Moon:2011sz}
\bibitem{Moon:2011sz}
  T.~Moon, Y.~S.~Myung and E.~J.~Son,
  %``Stability analysis of f(R)-AdS black holes,''
  Eur.\ Phys.\ J.\ C {\bf 71}, 1777 (2011)
  doi:10.1140/epjc/s10052-011-1777-0
  [arXiv:1104.1908 [gr-qc]].
  %%CITATION = doi:10.1140/epjc/s10052-011-1777-0;%%
  %20 citations counted in INSPIRE as of 10 Jan 2018


%\cite{Hirayama:2001bi}
\bibitem{Hirayama:2001bi}
  T.~Hirayama and G.~Kang,
  %``Stable black strings in anti-de Sitter space,''
  Phys.\ Rev.\ D {\bf 64}, 064010 (2001)
  doi:10.1103/PhysRevD.64.064010
  [hep-th/0104213].
  %%CITATION = doi:10.1103/PhysRevD.64.064010;%%
  %57 citations counted in INSPIRE as of 10 Jan 2018

%\cite{Eardley:1974nw}
\bibitem{Eardley:1974nw}
  D.~M.~Eardley, D.~L.~Lee and A.~P.~Lightman,
  %``Gravitational-wave observations as a tool for testing relativistic gravity,''
  Phys.\ Rev.\ D {\bf 8}, 3308 (1973).
  doi:10.1103/PhysRevD.8.3308
  %%CITATION = doi:10.1103/PhysRevD.8.3308;%%
  %105 citations counted in INSPIRE as of 21 Feb 2018


%\cite{Moon:2014qma}
\bibitem{Moon:2014qma}
  Y.~S.~Myung and T.~Moon,
  %``Massive gravitational waves in Chern-Simons modified gravity,''
  JCAP {\bf 1410}, no. 10, 043 (2014)
  doi:10.1088/1475-7516/2014/10/043
  [arXiv:1403.5433 [gr-qc]].
  %%CITATION = doi:10.1088/1475-7516/2014/10/043;%%
  %1 citations counted in INSPIRE as of 21 Feb 2018


%\cite{Lu:2011zk}
\bibitem{Lu:2011zk}
  H.~Lu and C.~N.~Pope,
  %``Critical Gravity in Four Dimensions,''
  Phys.\ Rev.\ Lett.\  {\bf 106}, 181302 (2011)
  doi:10.1103/PhysRevLett.106.181302
  [arXiv:1101.1971 [hep-th]].
  %%CITATION = doi:10.1103/PhysRevLett.106.181302;%%
  %166 citations counted in INSPIRE as of 10 Jan 2018


%\cite{Wald:1993nt}
\bibitem{Wald:1993nt}
  R.~M.~Wald,
  %``Black hole entropy is the Noether charge,''
  Phys.\ Rev.\ D {\bf 48}, no. 8, R3427 (1993)
  doi:10.1103/PhysRevD.48.R3427
  [gr-qc/9307038].
  %%CITATION = doi:10.1103/PhysRevD.48.R3427;%%
  %1218 citations counted in INSPIRE as of 02 Feb 2017

%\cite{Iyer:1994ys}
\bibitem{Iyer:1994ys}
  V.~Iyer and R.~M.~Wald,
  %``Some properties of Noether charge and a proposal for dynamical black hole entropy,''
  Phys.\ Rev.\ D {\bf 50}, 846 (1994)
  doi:10.1103/PhysRevD.50.846
  [gr-qc/9403028].
  %%CITATION = doi:10.1103/PhysRevD.50.846;%%
  %1016 citations counted in INSPIRE as of 02 Feb 2017

  %\cite{Jacobson:1993vj}
\bibitem{Jacobson:1993vj}
  T.~Jacobson, G.~Kang and R.~C.~Myers,
  %``On black hole entropy,''
  Phys.\ Rev.\ D {\bf 49}, 6587 (1994)
  doi:10.1103/PhysRevD.49.6587
  [gr-qc/9312023].
  %%CITATION = doi:10.1103/PhysRevD.49.6587;%%
  %446 citations counted in INSPIRE as of 02 Feb 2017

%\cite{Kwon:1986dw}
\bibitem{Kwon:1986dw}
  O.~J.~Kwon, Y.~D.~Kim, Y.~S.~Myung, B.~H.~Cho and Y.~J.~Park,
  %``Stability of the Schwarzschild black hole in Brans-Dicke theory,''
  Phys.\ Rev.\ D {\bf 34}, 333 (1986).
  doi:10.1103/PhysRevD.34.333
  %%CITATION = doi:10.1103/PhysRevD.34.333;%%
  %35 citations counted in INSPIRE as of 22 Feb 2018

%\cite{Hennigar:2018hza}
\bibitem{Hennigar:2018hza}
  R.~A.~Hennigar, M.~B.~J.~Poshteh and R.~B.~Mann,
  %``Shadows, Signals, and Stability in Einsteinian Cubic Gravity,''
  arXiv:1801.03223 [gr-qc].
  %%CITATION = ARXIV:1801.03223;%%


\end{thebibliography}
\end{document}